\documentclass{rnaastex}
\usepackage{amsmath}
\renewcommand{\vec}[1]{\boldsymbol{#1}}

\begin{document}

\title{\textit{NHDS}: The New Hampshire Dispersion relation Solver}

\correspondingauthor{Daniel Verscharen}
\email{d.verscharen@ucl.ac.uk}

\author[0000-0002-0497-1096]{Daniel Verscharen}
\affiliation{Mullard Space Science Laboratory, University College London, UK}
\affiliation{Space Science Center, University of New Hampshire, USA}

\author[0000-0003-4177-3328]{Benjamin D.~G.~Chandran}
\affiliation{Space Science Center, University of New Hampshire, USA}
\affiliation{Department of Physics, University of New Hampshire, USA}


\section{} 
In collisionless astrophysical plasmas, waves and instabilities are well modeled by the linearized Vlasov--Maxwell equations, which have non-trivial solutions only when the complex frequency $\omega$ solves the hot-plasma dispersion relation. \textit{NHDS} (New Hampshire Dispersion relation Solver) is a numerical tool written in Fortran 90 and first introduced by \citet{verscharen13} to solve this dispersion relation under the assumption that the plasma background distribution is a gyrotropic drifting bi-Maxwellian for each species $j$,
\begin{equation}
f_{0j}(v_{\perp},v_{\parallel})=\frac{n_j}{\pi^{3/2}w_{\perp j}^2w_{\parallel}}\exp\left(-\frac{v_{\perp}^2}{w_{\perp j}^2}-\frac{\left(v_{\parallel}-U_j\right)^2}{w_{\parallel j}^2}\right),
\end{equation}
in a cylindrical coordinate system aligned with the direction of the background magnetic field $\vec B_0$, where $n_j$ is the density, $w_{\perp}$ ($w_{\parallel}$) is the perpendicular (parallel) thermal speed with respect to $\vec B_0$, and $U_j$ is the field-aligned drift speed. All floating-point quantities use double precision. 

The \textit{NHDS} code closely follows the formulation of the hot-plasma dispersion relation laid out by \citet{stix92}. It uses a Newton-secant method to identify those frequencies at which there are non-trivial solutions to the wave equation,
\begin{equation}\label{waveeq}
\begin{pmatrix}\displaystyle\epsilon_{xx}-\frac{k_z^2c^2}{\omega^2} & \epsilon_{xy} & \displaystyle\epsilon_{xz}+\frac{k_{\perp}k_zc^2}{\omega^2} \\ \epsilon_{yx} & \displaystyle\epsilon_{yy}-\frac{k^2c^2}{\omega^2} & \epsilon_{yz} \\ \displaystyle\epsilon_{zx}+\frac{k_{\perp}k_zc^2}{\omega^2} & \epsilon_{zy} & \displaystyle\epsilon_{zz}-\frac{k_{\perp}^2c^2}{\omega^2}\end{pmatrix} \begin{pmatrix}E_x \\ E_y \\ E_z\end{pmatrix}=0,
\end{equation}
based on an initial guess for $\omega$, where $\epsilon$ is the dielectric tensor, $\vec E$ is the vector of the electric-field Fourier amplitudes, $c$ is the speed of light, and $\vec k=(k_{\perp},0,k_z)$ is the wavevector. The initial guess defines the plasma mode that the code follows in $\vec k$. The Newton-secant method converges if the absolute value of the determinant of the matrix in Equation~(\ref{waveeq}) is less than a user-defined value.
All frequencies are given in units of the proton gyro-frequency $\Omega_{\mathrm p}$ and all length scales in units of the proton inertial length $d_{\mathrm p}$.

For each of the up to ten plasma components $j$, the user defines the temperature anisotropy $T_{\perp j}/T_{\parallel j}$ with respect to $\vec B_0$, the value of $\beta_{\parallel j}\equiv 8\pi n_jk_{\mathrm B}T_{\parallel j}/B_0^2$, the relative charge $q_j/q_{\mathrm p}$, the relative mass $m_j/m_{\mathrm p}$, the relative density $n_j/n_{\mathrm p}$, and the normalized drift velocity $U_j/v_{\mathrm A}$, where $k_{\mathrm B}$ is the Boltzmann constant, $v_{\mathrm A}$ is the proton Alfv\'en speed, and $T_{\parallel j}$ is the temperature parallel to $\vec B_0$. Furthermore, the ratio $v_{\mathrm A}/c$ and the angle of propagation $\theta$ are user-defined parameters.

The calculation of $\epsilon_{ik}$ entails the evaluation of the modified Bessel function $I_m(\lambda_j)$ of the first kind and the plasma dispersion function $Z(\zeta)$, where $\lambda_j\equiv k_{\perp}^2w_{\perp j}^2/2\Omega_j^2$, and $\zeta$ is a dimensionless complex number. For the evaluation of $I_m$, \textit{NHDS} applies the recursion method supplied by the \textit{Numath} Library \citep{clenshaw62}. It determines the maximum order $m_{\max}$ of $I_m$ as the smaller  of either a user-defined limit or as the number for which $I_{m_{\max}}(\lambda_j)$ is less than a user-defined value. \textit{NHDS} evaluates $Z(\zeta)$ following \citet{poppe90} by computing the complex error function $w(\zeta)=Z(\zeta)/i\sqrt{\pi}$ through one of the following methods, depending on the value of $|\zeta|$: a power series, the Laplace continued fraction method, or a truncated Taylor expansion. This combined method is faster than alternative approaches and calculates $w(\zeta)$ to an accuracy of 14 significant digits for almost all $\zeta$. 

\textit{NHDS} determines the polarization of the wave solutions as the ratios $E_y/E_x$ and $E_z/E_x$ from Equation~(\ref{waveeq}), which translate to ratios of the magnetic-field amplitudes through Faraday's law. 
In addition, as described by \citet{verscharen13a} and \citet{verscharen16}, \textit{NHDS} calculates the relative wave energy $W_k$ and the Fourier amplitudes of the fluctuations in density, bulk velocity, and pressure. The code also calculates the contribution $\gamma_j$ to the total growth/damping rate $\mathrm{Im}(\omega)$ from each species $j$ as described by \citet{quataert98}. 

For a given wave solution, \textit{NHDS} can determine the value of the self-consistent fluctuating distribution function on a user-defined Cartesian grid in velocity space as described by \citet{verscharen16}.  \textit{NHDS} saves the fluctuating distribution function in  \textit{HDF5} files and creates an \textit{XDMF} file for visualization with programs like \textit{ParaView}. This calculation entails the calculation of the Bessel function $J_m(k_{\perp}v_{\perp}/\Omega_j)$ of order $m$, which \textit{NHDS} performs through a polynomial Chebyshev approximation. The maximum order $m_{\max}$ for $J_m$ is determined in the same way as $m_{\max}$ for $I_m$ in the calculation of $\epsilon_{ik}$, except that $m_{\max}$ for $J_m$ is evaluated for each $v_{\perp}$.

\begin{figure}[h!]
\begin{center}
\includegraphics[scale=0.85,angle=0]{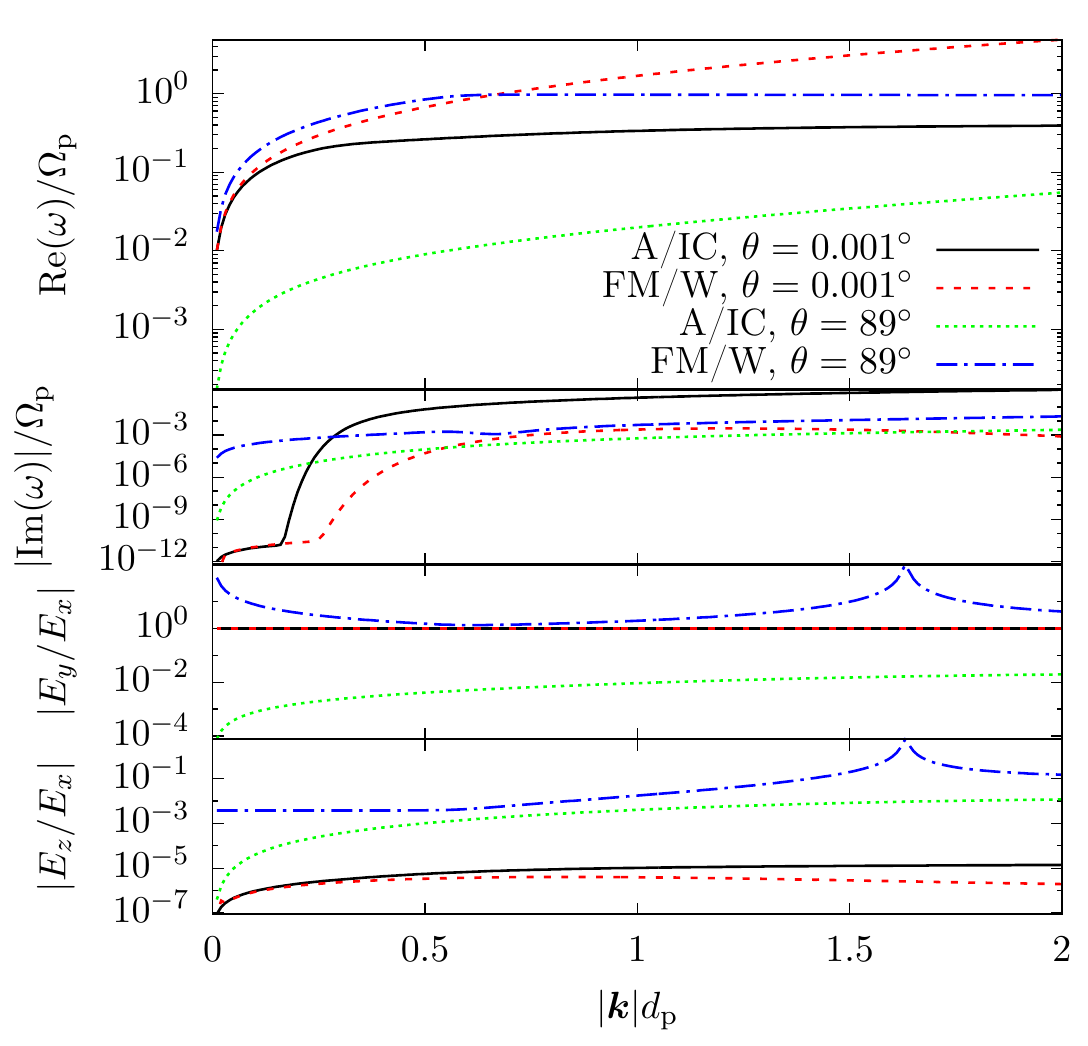}
\caption{Dispersion relations for the A/IC and FM/W waves in parallel ($\theta=0.001^{\circ}$) and perpendicular ($\theta=89^{\circ}$) propagation. The panels show from the top to the bottom: the normalized real part of the frequency, the normalized damping rate, the ratio $|E_y/E_x|$, and the ratio $|E_z/E_x|$ as functions of $|\vec k|$. \label{figure}}
\end{center}
\end{figure}

Figure~\ref{figure} shows the dispersion relations of Alfv\'en/ion-cyclotron (A/IC) and fast-magnetosonic/whistler (FM/W) waves in parallel and perpendicular propagation as well as some of their polarization properties determined with \textit{NHDS}.

The code is publicly available for download \citep[][Codebase: \url{https://github.com/danielver02/NHDS}]{verscharen18}.

\acknowledgments

We acknowledge support from NASA, NSF, and STFC. We appreciate discussions with Kris Klein.

\end{document}